\documentstyle[prd,aps,floats,psfig]{revtex}

\newcommand{\pabar}{\not{\!\partial}}
\newcommand{\Od}{{\cal O}}

\newcommand{\Dbar}{\not{\!{\!D}}}

\newcommand{\re}{\mbox{Re}}

\def\gappeq{\mathrel{\rlap {\raise.5ex\hbox{$>$}}
{\lower.5ex\hbox{$\sim$}}}}

\def\lappeq{\mathrel{\rlap{\raise.5ex\hbox{$<$}}
{\lower.5ex\hbox{$\sim$}}}}
\begin{document}
\draft
\input epsf
\renewcommand{\topfraction}{0.8}
\preprint{CERN-TH/99-373}
\title{The equivalence theorem and the production of gravitinos after 
inflation}

\author{Antonio L. Maroto\footnote{E-mail: Antonio.Lopez.Maroto@cern.ch}}
\address{CERN Theory Division, \\
CH-1211 Geneva 23, Switzerland}
\author{J. R. Pel\'aez  \footnote{E-mail:pelaez@eucmax.sim.ucm.es}}
\address{ Departamento de F\'{\i}sica Te\'orica. \\
Universidad Complutense. 28040 Madrid. SPAIN.}
\date{November 1999} 
\maketitle
We study the high-energy equivalence between 
helicity $\pm 1/2$ gravitinos
and goldstinos in order to calculate 
the production of gravitinos in time-dependent
scalar and gravitational backgrounds. We derive this equivalence for 
equations of motion, 
paying attention to some subtleties, mainly
due to external sources, that are not present
in the standard proofs.
We also propose the Landau gauge as a simplifying 
alternative to the usual gauge choices, both for practical calculations
and in the equivalence theorem proof. 
\begin{abstract}

\end{abstract}
%

\section{Introduction}
In supergravity theories \cite{Nilles,Bailin} 
 the graviton superpartner is a 
spin $3/2$ particle called the gravitino. This particle couples only
with gravitational strength to the rest of matter fields, and
accordingly its lifetime can be very long, with a decay rate of 
$\Gamma_{3/2}\simeq m_{3/2}^3/M_P^2$. 
In particular,  gravitinos lighter than $m_{3/2}<100\,\hbox{MeV}$ will 
live longer
than the age of the Universe. 
This fact can have important consequences in cosmology
 and imposes stringent constraints on supergravity models.
Owing to their weak couplings, gravitinos
freeze out very early when they are still relativistic;  
therefore their primordial abundance can be estimated as 
$n_{3/2}/s\simeq 10^{-3}$
\cite{Kolb}.
Considering only the case of unstable gravitinos, such a
primordial abundance
would give rise to an enormous amount of entropy, in conflict 
with the standard
cosmology. In particular, gravitinos decaying during the
nucleosynthesis
can destroy the nuclei created in that era. A possible way out of 
this {\it gravitino problem}, 
is the existence of a period of inflation that dilutes any 
primordial density \cite{LiEll}.
Unfortunately the problem can be re-created if, after inflation,
gravitinos 
are produced
by some mechanism. In fact, this could be the case if during the 
period of inflaton oscillations, at the end of inflation, 
the reheating  temperature
was sufficiently high.
A successful nucleosynthesis era then requires (we give some conservative
bounds \cite{Sarkarrep}): 
$n_{3/2}/s<10^{-15}$ for a gravitino  mass $m_{3/2}\simeq 100\,\hbox{GeV}$,
$n_{3/2}/s<10^{-14}$ for $m_{3/2}\simeq 1\,\hbox{TeV}$ and $n_{3/2}/s<10^{-13}$
for $m_{3/2}\simeq 10\,\hbox{TeV}$. 
The production of gravitinos during reheating is due to
processes involving other particles produced from the inflaton decay,
and depends on the reheating temperature $T_R$ as \cite{Ellis}:
$n_{3/2}/s\simeq 10^{-14}T_R/(10^9\; \hbox{GeV})$.
For a typical  mass $m_{3/2}\simeq 1 \,\hbox{TeV}$, 
this implies $T_R<10^9\; \hbox{GeV}$.
Another constraint appears in supergravity models where
the gravitino mass is determined by the scale of 
supersymmetry breaking. In order to solve the hierarchy problem, it is 
then suggested 
that $m_{3/2}<1 \,\hbox{TeV}$ \cite{Bailin}.

However, as a consequence of some recent works  \cite{LindeB} 
it was realized that,  
during the first inflaton oscillations, reheating
can not be studied by the standard perturbative
techniques.
This  {\it preheating} period can give rise to an explosive production
of bosons due to the phenomenon of parametric resonance. 
In this period, the energy
of the coherent oscillations of the  inflaton field is very efficiently 
converted into particles.  In the
case of fermions, the limit imposed by the Pauli exclusion principle avoids the
explosive production, 
although the results still deviate from the perturbative
expectations \cite{Baacke,Green}. 
This fact is particularly relevant when gravitinos
are directly coupled to the inflaton, since
during preheating they could be produced in excess,
thus imposing new constraints on the particular supergravity
inflationary model. 

In  previous works \cite{Maroto,Valencia} it was shown  that the 
production of helicity $\pm 3/2$ gravitinos can take place during 
preheating and that the results deviate from the perturbative 
expectations by several orders of magnitude (see also \cite{Lemoine}).
In the case of helicity $\pm 1/2$ gravitinos, the production is in general 
more abundant, depending on the specific supergravity model
\cite{Linde}. Some other works dealing with this topic can be found in 
\cite{varios}. 
 
In the present work we are interested in the production of 
helicity $\pm 1/2$ gravitinos during preheating. 
The relative difficulty of the calculations in the unitary gauge,  
used in the above references, suggests that we should explore 
alternative 
methods. In particular we will exploit the 
relation between helicity $\pm 1/2$ gravitinos and goldstinos
(first pointed out in \cite{Fayet})
given by the Equivalence Theorem (ET) \cite{Casalbuoni}.
This possibility was suggested in the first reference of 
 \cite{Linde} and
in the last one of \cite{Linde}
it was shown how the ET could be used 
to study the helicity $\pm1/2$ gravitino equation in
the $\vert \phi \vert \ll M_P$ limit.

The ET was first 
introduced, in the framework of non-abelian gauge theories
\cite{Cornwall,ETgauge}, as a way
to calculate processes involving longitudinal gauge bosons, but using
only Goldstone bosons, which are scalars and therefore much easier to handle.
The first formal proof in terms of S-matrix elements was given in 
\cite{Chanowitz},
and that is basically the derivation followed
within the supergravity scenario. In the last few years, and still within
the framework of the non-abelian gauge theory, several works
have completed the proof of the theorem \cite{ETrenormalizado},
including renormalization effects, 
but also raising some questions about its Lorentz non-invariance ambiguity and 
applicability \cite{ETLorentz}. In this paper we  also discuss briefly
how these new considerations may affect the gravitino-goldstino 
ET 
applied to the production of gravitinos during preheating.

Intuitively, the ET tells us that, since 
the goldstinos disappear from the spectrum through the super-Higgs
mechanism, giving rise
to physical helicity $\pm 1/2$ gravitinos, it is possible to
use goldstinos in the calculation of observables instead of the 
complicated $\pm 1/2$ gravitinos. Of course, this identification can only
be carried out at energies high enough to neglect the masses.

Rigorously, this theorem has only been proved for S-matrix elements
containing initial or final helicity $\pm 1/2$ gravitinos
and in the absence of external backgrounds. This would provide
a good approximation for gravitino production during the reheating period,
but only at the perturbative level, where the rate of production
is given by the decay of inflaton quanta \cite{LindeB,Casalbuoni}.

However, preheating is 
a non-perturbative (and out of equilibrium) 
process and it is not obvious that the same proof 
still holds in the presence of
external sources, such as the inflaton field or the space-time
curvature. In particular, the presence
of a source that creates particles makes different the initial and final
vacua in the Green functions. 
In addition, these sources are present in the
gauge-fixing condition, which is the starting relation 
in the ET derivations.  
Of course, we still expect that the intuitive relation suggested by
the ET should hold, but since it is not the same to establish an
equality  at the level of matrix elements as at the operator 
(fields, indeed) level, 
we present in this paper a derivation more suited for the formalism
in terms of equations of motion. In this way we can also identify the
physical conditions on the sources that we need for this theorem to
hold. 

Finally, we propose the Landau gauge as
the best choice to perform the calculations, although, probably, it is
not the most intuitive. In this gauge, not only the proof of the 
theorem, but
also the final equations that govern gravitino production
are considerably simpler.

All the previous considerations basically concern the 
gravitino production
process. But we also have to take into account the fact 
that we are producing very many gravitinos (out of equilibrium)
which have a distribution in energies. Some of them will satisfy the
physical conditions to apply the ET, whereas some others will not.
Hence, we also present an additional condition 
on the number of those gravitinos not satisfying the 
applicability conditions, in order to obtain reliable
calculations with the ET.

\section{Supergravity Lagrangian}
Let us consider $N=1$ minimal supergravity \cite{Nilles,Bailin,Julia}
 coupled to a single chiral superfield 
$\Phi$, 
which describes a complex scalar field $\phi$ and a Majorana spinor
$\eta$ satisfying $\eta=C\bar\eta^T=\eta^C$, with the
charge-conjugation matrix given by  $C=i\gamma^2\gamma^0$.
In principle, the derivation could be extended to more than one chiral 
multiplet in a similar way.  
The scalar component will  play the role of the inflaton field and
it will therefore be considered as an external background.
The corresponding Lagrangian is defined by the superpotential
$W(\Phi)$ and the  K\"ahler
potential $G(\Phi,\Phi^\dagger)=\Phi^\dagger\Phi+\log \vert W \vert^2$.
We will define: $G_{,\phi}=\partial G/\partial \phi$,
$G_{,\phi^*}=\partial G/\partial \phi^*$, 
$G_{,\phi\phi^*}=\partial^2 G/\partial \phi \partial \phi^*$, etc. 
In this case we will have $G_{,\phi\phi^*}=1$. 
The bosonic part of the Lagrangian is given by
\begin{eqnarray}
g^{-1/2}{\cal L}_B=-\frac{1}{2}R+g^{\mu\nu}
\partial_\mu \phi \partial_\nu \phi^*+e^{G}\left(3
-\vert G_{,\phi}\vert ^2\right),
\end{eqnarray}
where we are working in units $M=M_P/\sqrt{8\pi}=1$. In the fermionic
part of the Lagrangian, we are only interested in those terms
quadratic
in the fermionic fields (gravitinos and goldstinos), 
since we are going to work with the
linearized equations of motion. For the sake of simplicity
we will assume that the scalar field $\phi$ is real. With this
assumption those terms are:
\begin{eqnarray}
g^{-1/2}{\cal L}_F&=&-\frac{1}{2} \epsilon^{\mu\nu\rho\sigma}
\bar\psi_\mu \gamma_5 \gamma_\nu D_\rho
\psi_\sigma+\frac{i}{2}\bar \eta
\Dbar \eta +e^{G/2}\left( \frac{i}{2}\bar \psi^\mu \sigma_{\mu\nu}
\psi^\nu+\frac{1}{2}\left(-G_{,\phi\phi}-G_{,\phi}^2\right)
\bar \eta \eta \right. \nonumber \\
&+&\left. \frac{i}{\sqrt{2}}G_{,\phi}\bar \psi_\mu
\gamma^\mu\eta\right)+\frac{1}{\sqrt{2}}\bar \psi_\mu 
(\pabar \phi) \gamma^\mu
\eta,
\label{lagrangiano}
\end{eqnarray}
with $\sigma_{\mu\nu}=\frac{i}{2}[\gamma_\mu,\gamma_\nu]$.
Since we  are concerned with the production of gravitinos
after inflation, we assume that our scalar field
depends only on time and that the space-time metric is of the
Friedmann-Robertson-Walker (FRW) form. In particular, it will be very
useful to work in conformal time, for which the FRW metric with
flat spatial sections reads:
\begin{eqnarray}
ds^2=a^2(t)(dt^2-d\vec x^2),
\end{eqnarray}
where $a(t)$ is the Universe scale factor and the non-vanishing
gravitational
field is assumed to be created by the scalar field.

In contrast with the ET usual proof, 
there are two mixing terms between gravitinos and goldstinos 
in eq. (\ref{lagrangiano}).
When the scalar field has settled down at the potential minimum, 
$\phi=\phi_0$, the last term does not contribute, 
and this is
why it is absent from the discussions of the 
spontaneous breaking of supersymmetry. However, since we are interested
in a time-dependent $\phi$, such a term cannot be
ignored any longer.
In flat space-time, with $\phi=\phi_0$, and when supersymmetry is not 
broken, i.e. $G_{,\phi_0}=0$,
the mixing terms are absent and the equations of motion describe the
gravitino evolution with only
two helicity $\pm 3/2$ states. However,  when supersymmetry is
broken spontaneously,  the gravitino
acquires two more degrees of freedom with helicity $\pm 1/2$,
because of the interaction with the goldstinos,
giving rise to  much more complicated evolution
equations. 

In the unitary gauge all the goldstino dependent terms
are absorbed in a redefinition of the gravitino field.
This gauge shows explicitly the super-Higgs mechanism in which
the goldstino becomes the helicity $\pm 1/2$ components of the gravitino
field. There are no mixing terms but, still, we have to deal 
with $\pm 1/2$ helicity
states of a Rarita-Schwinger field, which can be rather involved.
Nevertheless,
the production of helicity $\pm 1/2$ gravitinos in preheating
has been calculated in the unitary gauge in \cite{Linde}.

Note that, although it is not necessary,
we are making the inflaton
responsible for supersymmetry breaking.
Then, the
inflatino also plays the role of the goldstino.
This assumption simplifies the discussion since
otherwise, and although supersymmetry would be broken 
during and after
inflation, it would be restored at the minimum of the potential and
the super-Higgs mechanism would not take place.
Accordingly, the gravitino would not have a $\pm 1/2$ component. 
We will also assume that at the minimum the cosmological 
constant is zero, then $G_{,\phi_0}^2=3G_{,\phi_0\phi_0^*}$, and we have
$G_{,\phi_0}=\sqrt{3}$.

The equations of motion for gravitinos and goldstinos derived from
eq. (\ref{lagrangiano}) are
\begin{eqnarray}
\epsilon^{\mu\nu\rho\sigma}\gamma_5\gamma_\nu D_\rho\psi_\sigma+
\frac{1}{2}e^{G/2}[\gamma^\mu,\gamma^\nu]\psi_\nu
-\frac{i}{\sqrt{2}}G_{,\phi}e^{G/2}\gamma^\mu
\eta-\frac{1}{\sqrt{2}}
(\pabar \phi)\gamma^\mu\eta=0
\label{eomgravitino}
\end{eqnarray}
and
\begin{eqnarray}
i\Dbar \eta + e^{G/2}\left(-G_{,\phi\phi}
-G_{,\phi}^2\right)\eta
-\frac{i}{\sqrt{2}}e^{G/2}G_{,\phi}
\gamma^\mu\psi_\mu+\frac{1}{\sqrt{2}}\gamma^\mu(\pabar
\phi)
\psi_\mu=0
\label{gold0}
\end{eqnarray}
If we consider only  helicity $\pm 3/2$ gravitinos, then it can be shown
that the equations of motion  reduce to a very simple form
\cite{Maroto}:
\begin{eqnarray}
(i\Dbar-e^{G/2})\psi_\mu^{\pm 3/2}=0
\label{g32}
\end{eqnarray}
However, the helicity $\pm1/2$ equation
 is much more involved and contains terms
coupled to goldstinos.
Nevertheless, if we are only interested in the helicity $\pm 1/2$ gravitinos
high-energy behavior, $E \gg m_{3/2}$, then we can
 simplify the calculations with the ET.
This limit is sensible in most supergravity inflation models
with one chiral supermultiplet,
since the typical energy of the
particles created during preheating is of the order of the inflaton mass,
$m_{\phi}$,
which is usually several orders of magnitude larger than $m_{3/2}$. 
For instance, in the model discussed in 
\cite{Holman,Ross}, $m_{\phi}\simeq
10^{10}\,\hbox{GeV}$, 
whereas $m_{3/2}< 1 \,\hbox{TeV}$. Note also
that all these scales are well below $M_P$, where supergravity
breaks down as an effective theory.

As it was commented before, the ET has been
rigorously derived for S-matrix elements. However,
to calculate the non-perturbative production of gravitinos
during preheating, we use a formalism
in terms of equations of motion and fields.
In order to show how the gravitino-goldstino high-energy equivalence can be
used in this context, we will follow these steps:

i) Introduce a gauge-fixing term corresponding to a certain
generalization of the $R_{\xi}$ gauges, which allows us to cancel
the mixing gravitino-goldstino terms in the equations of motion.

ii) Assume that in  the asymptotic regions $t\rightarrow \pm \infty$,
the external sources are static, i.e, $\phi\rightarrow \phi_0$  and 
$g_{\mu\nu}\rightarrow \eta_{\mu\nu}$, and then use the equations
of motion in those regions to show that 
$\partial^{\mu}\psi_\mu\propto m_{3/2}\eta$.

iii) Use the high-energy limit of the $\pm1/2$ helicity projectors,
$P^\mu_{\pm1/2}=p^\mu/m_{3/2}+\Od(m_{3/2}/E)$, to relate
$\psi_{\pm1/2}=P^\mu_{\pm 1/2}\psi_\mu\propto\eta$, when $E\gg m_{3/2}$
in the asymptotic regions.

iv) Choose the Landau gauge, $\xi \rightarrow \infty$, as
an additional simplification for the calculations
of goldstino production.
     
\section{Gauge fixing}

Goldstinos
do not belong to the physical spectrum, and in the unitary
gauge we can even get rid of them in the equations of motion.
In contrast, 
the production of helicity $\pm1/2$ gravitinos during
reheating is gauge-invariant, and is only related
to the goldstino production in certain  gauges, called $R_\xi$ gauges, 
in which both fields appear simultaneously in the Lagrangian.
Let us then consider the following gauge-fixing condition, which is a
generalization
of the $R_\xi$ gauge used in  \cite{Casalbuoni,Baulieu}:
\begin{eqnarray}
\gamma^\mu\psi_\mu
-\frac{1}{\sqrt{2}\xi\Dbar}e^{G/2}G_{,\phi}\eta
+\frac{i}{G_{,\phi}}e^{-G/2}\gamma^\mu(\pabar\phi)\psi_{\mu}=0.
\end{eqnarray}
When $\phi$ is constant we recover the gauge-fixing term in
\cite{Casalbuoni} and the limit
$\xi\rightarrow 0$ corresponds to the unitary gauge. 
Note that in our case, {\it due to the
 external sources}, all the coefficients in the
gauge-fixing function are no longer constants.
The above equation provides us with a relation
between gravitinos and goldstinos, but we want to
extract only those with  helicity $\pm1/2$,
for which we will need a relation between $\partial^\mu \psi_\mu$
and $\eta$. In the following we will use the equations
of motion to obtain a relation of the desired form.

If we assume that in the asymptotic regions $t\rightarrow \pm\infty$ the
space-time is flat and the scalar field settles down at the 
potential minimum $\phi_0$, then, in those regions,
the above condition reduces to
\begin{eqnarray}
a^{-1}_{in,out}\pabar\gamma^\mu\psi_\mu
=\sqrt{\frac{3}{2}}\frac{m_{3/2}}{\xi}\eta,
\end{eqnarray}
where $m_{3/2}=e^{G_0/2}$ and $a_{in,out}$ are the scale factor
values in the asymptotic past and future. In order to simplify the notation,
we will absorb the scale factor into the mass: $m_{in,out}\equiv
a_{in,out} m_{3/2}$; to avoid the
proliferation of indices, we will denote $m_{in,out}$ simply by $m$.
With this notation, the gauge-fixing condition reads:
\begin{eqnarray}
\pabar\gamma^\mu\psi_\mu
=\frac{m}{\xi}\sqrt{\frac{3}{2}}\eta.
\label{gfsimple}
\end{eqnarray}

Let us recall that it is only in the static regions
where the definition of particle and the separation between different
helicities  is unambiguous. However, in the strict sense,
within the inflationary cosmology neither the initial
($t\rightarrow -\infty$) nor the final ($t\rightarrow
\infty$) regions can be considered  static, since 
there is a  period of inflation before preheating  
and today we know that the Universe is expanding. Nevertheless, 
for practical purposes, we can still consider
the initial and final regions as static, since the 
particle production will mainly take place during the first inflaton
oscillations. Accordingly, 
we will define our initial vacuum by imposing such  initial
conditions on our fields  that they behave as plane
waves before preheating. 
The final state has a similar behavior, since the rate of expansion
decreases with time.  Indeed, the vacuum at the end of preheating 
could be defined more rigorously as an adiabatic vacuum
\cite{Birrell}, which would not yield  additional
gravitino production from the Universe expansion. 

Let us  then consider first the equations of motion for gravitinos 
eq. (\ref{eomgravitino}) and
goldstinos eq. (\ref{gold0}) in the initial and 
final regions with the notation
that we have just introduced. Since the inflaton is in the minimum, 
$G_{,\phi_0^*}G_{,\phi_0\phi_0}=-G_{,\phi_0}$, and therefore
\begin{eqnarray}
&&\epsilon^{\mu\nu\rho\sigma}\gamma_5\gamma_\nu \partial_\rho\psi_\sigma+
\frac{1}{2}m[\gamma^\mu,\gamma^\nu]\psi_\nu
-i\sqrt{\frac{2}{3}}m\gamma^\mu
\eta=0,\\
&&i\pabar \eta -2m\eta
-i\sqrt{\frac{3}{2}}m\gamma^\mu\psi_\mu=0.
\end{eqnarray}
If we now fix the gauge using eq.(\ref{gfsimple}) in the above
equations, they can be rewritten as
\begin{eqnarray}
&&\epsilon^{\mu\nu\rho\sigma}\gamma_5\gamma_\nu \partial_\rho\psi_\sigma+
\frac{1}{2}m[\gamma^\mu,\gamma^\nu]\psi_\nu
-i\xi \gamma^\mu \pabar \gamma^\nu \psi_\nu=0,\\
&&i\pabar \eta -2m\eta
-i\frac{3}{2}\frac{m^2}{\xi}\frac{1}{\pabar}\eta=0,
\label{gold}
\end{eqnarray}

In the following we will rewrite the equations of motion for goldstinos
and gravitinos in the asymptotic regions
as well as the gauge-fixing condition in a more convenient form.
Contracting the gravitino equation with $\partial_\mu$,
 we obtain \cite{Moroi}
\begin{eqnarray}
\frac{1}{2}m(\pabar\gamma^\nu\psi_\nu-\gamma^\nu\pabar\psi_\nu)
-i\xi\pabar\pabar \gamma^\nu \psi_\nu=0,
\label{pri}
\end{eqnarray}
whereas contracting with $\gamma_\lambda \gamma_\mu$, we find
\begin{eqnarray}
2i(\partial_\lambda\gamma^\sigma\psi_\sigma-\pabar\psi_\lambda)
+m(\gamma_\lambda\gamma^\nu\psi_\nu+2\psi_\lambda)-2i\xi\gamma_\lambda
\pabar\gamma^\nu\psi_\nu=0,
\end{eqnarray}
which can be contracted again with $\gamma_\lambda$ to get
\begin{eqnarray}
i(\pabar\gamma^\sigma\psi_\sigma-\gamma^\lambda\pabar\psi_\lambda)+
3m\gamma^\mu\psi_\mu-4i\xi \pabar\gamma^\nu\psi_\nu=0.
\label{seg}
\end{eqnarray}
Substituting eq. (\ref{pri}) into eq. (\ref{seg}), we obtain
\begin{eqnarray}
-i\xi \pabar\pabar
 \gamma^\nu\psi_\nu+\frac{3}{2}im^2\gamma^\nu\psi_\nu
+2\xi m\pabar \gamma^\nu \psi_\nu=0,
\end{eqnarray}
which, finally, can be rewritten as:
\begin{eqnarray}
(i\pabar -m_+)(i\pabar-m_-)\gamma^\nu\psi_\nu=0,
\label{eomparagammapsi}
\end{eqnarray}
where we have defined $m_{\pm}=m(1\pm \sqrt{1-3/(2\xi)})$. 
Note that, in the perturbative sense,
the poles in the propagator are exactly those obtained in \cite{Casalbuoni}.
In addition we can derive the very same equation for the
goldstino, just by multiplying eq. (\ref{gold}) by $i\pabar$:
\begin{eqnarray}
(i\pabar -m_+)(i\pabar-m_-)\eta=0
\label{gold2}
\end{eqnarray}
The implications of these equations are clearer 
if we recall that the physical fields, i.e. 
the $\pm 3/2$ 
and $\pm 1/2$ helicity modes in the asymptotic regions,
 are those satisfying both $\gamma^\mu\psi_\mu=0$ and
$\partial^\mu\psi_\mu=0$. 
They have the correct  physical mass $m$ since 
they still satisfy $(i\pabar-m)\psi_\mu^{Phy}=0$. 
In contrast, the unphysical spin-1/2 
modes present poles in the propagator at $m_{\pm}$, 
exactly as  happens with the goldstinos.
Therefore, by fixing the gauge we have only modified the poles of the
unphysical modes.
From (\ref{pri}) and by means of the gauge-fixing condition, we obtain
\begin{eqnarray}
\frac{1}{2}m(2\pabar\gamma^\nu\psi_\nu-2\partial^\nu\psi_\nu)
-i\pabar\sqrt{\frac{3}{2}}m\eta=0
\label{pre}
\end{eqnarray}
Since $\eta$ satisfies eq. (\ref{gold2}), we have two possible
solutions: $(i\pabar -m_+)\eta=0$
and $(i\pabar-m_-)\eta=0$; together with the gauge-fixing 
condition (\ref{pre}), they 
yield
\begin{eqnarray}
\sqrt{\frac{3}{2}}\frac{m}{\xi}\eta
-\partial^\mu\psi_\mu-\sqrt{\frac{3}{2}}
m_{\pm}\eta=0.
\end{eqnarray}
From this expression we get:
\begin{eqnarray}
\partial^\mu\psi_\mu=\sqrt{\frac{3}{2}}\frac{m}{\xi}
\left(1-\xi\frac{m_{\pm}}
{m}\right)
\eta.
\label{gg}
\end{eqnarray}
At first sight, this equation relates the unphysical gravitino 
$\partial^\mu\psi_\mu$ with the goldstino; however, the key
observation is that, as we will show,  at high energy, 
$\partial^\mu\psi_\mu$ tends to the physical helicity $\pm 1/2$ gravitino.
Note that, apparently, there are two relations, one for
goldstinos that correspond to the $m_-$ solution and
another for those with $m_+$.

\section{The equivalence theorem}

In the asymptotic initial and final regions, we expect that a general 
solution of the
equations of motion for gravitinos and goldstinos will be written as a
linear superposition of on-shell positive and negative frequency plane
waves \cite{Moroi}. 
In particular, let us consider a negative frequency mode solution 
in the initial region with momentum  $p^{\mu}=(\omega_{in},0,0,p)$
(there is no loss of generality in this choice),  
with $p_\mu p^\mu=m_{in}^2$
and $p=\vert \vec p \vert$, such
that $p\gg m_{in}$: 
\begin{eqnarray}
\psi_\mu^{p}(x)&=&\frac{1}{a^{3/2}_{in}\sqrt{2\omega_{in}}}
e^{i px}\tilde\psi_\mu(\vec p)
+\Od\left(\frac{m_{in}}{p}\right),
\label{grv}
\end{eqnarray}
where $\tilde \psi_\mu(\vec p)$ is the corresponding Fourier component.
 For the unphysical  goldstino we have
\begin{eqnarray}
\eta^{p}(x)&=&\frac{1}{a^{3/2}_{in}\sqrt{2\omega_{in}}}e^{i px} 
\tilde \eta(\vec p)
+\Od\left(\frac{m_{in}}{p}\right).
\label{goldi}
\end{eqnarray} 
Similar expressions can be written for the positive-frequency
solutions
and for solutions in the final region.
Note that the on-shell conditions for gravitinos and goldstinos
are different because of the different positions of the poles.
In particular, for physical gravitinos we have $p_\mu p^\mu=m_{in}^2$ and
for goldstinos and unphysical gravitinos ($\gamma^\mu\psi_\mu$ and 
$\partial^\mu\psi_\mu$) we have $p_\mu p^\mu=(m_{+}^{in})^2$ (we 
will not use $m_-$ for reasons that will become clear below). 
We have thus  
included the $\Od(m_{in}/p)$ term at the
end of (\ref{grv}) and (\ref{goldi}). Strictly, it should not
be there for physical gravitinos, but at this level we keep a 
compact notation between physical and unphysical gravitinos.


The spin-1 polarization vectors are given by 
$\epsilon_\mu(\vec p,m)=a(t)\delta^a_\mu \epsilon_a(\vec p,m)$, where
\begin{eqnarray}
\epsilon_a(\vec p, 1)= \frac{1}{\sqrt{2}} 
(0,1,i,0)\;\;, 
\epsilon_a(\vec p, 0)=\frac{1}{m_{in}}(p,0,0,\omega_{in})\;\;,
\epsilon_a(\vec p,-1)=-\frac{1}{\sqrt{2}}
(0,1,-i,0).
\end{eqnarray}
If $u(\vec p,s)$ are spinors with definite 
helicity $s=\pm 1/2$, then 
$P_\pm u(\vec p,\pm 1/2)= u(\vec p,\pm 1/2)$,  where 
$P_\pm=(1/2)(1\pm \gamma_5 \gamma^\mu\epsilon_\mu(\vec p, 0))$ 
are the helicity projectors. 
Accordingly, the helicity $\pm 3/2$ and $\pm 1/2$ projectors are nothing 
but
\begin{eqnarray}
P_\mu^{\pm 3/2}&=&P_{\pm}\epsilon_\mu(\vec p, \pm 1),\nonumber \\
P_\mu^{\pm 1/2}&=&\sqrt{\frac{1}{3}}P_{\mp}\epsilon_\mu(\vec p,\pm 1)
+\sqrt{\frac{2}{3}}P_{\pm}\epsilon_\mu(\vec p, 0).
\end{eqnarray}
We see that, {\em at high energy}, the $\pm 1/2$ projector behaves as
\begin{eqnarray}
P_\mu^{\pm 1/2}=\sqrt{\frac{2}{3}}P_{\pm}\frac{p_\mu}{m_{in}}
+\Od\left(\frac{m_{in}}{p}\right),
\end{eqnarray}
where we have neglected $\epsilon_\mu(\vec p, \pm 1)$
 with respect to $\epsilon_\mu(\vec p, 0)$. 
Let us then define the helicity $\pm 1/2$ components of the gravitino field
in the asymptotic initial regions and in momentum space as
\begin{eqnarray}
\tilde\psi_{\pm 1/2}(\vec p) \equiv P^\mu_{\pm
  1/2}\tilde\psi_\mu(\vec p)=\left[\sqrt{\frac{2}{3}}
P_{\pm}\frac{p^\mu}{m_{in}}
+\Od\left(\frac{m_{in}}{p}\right)\right]\tilde\psi_\mu(\vec p).
\label{preTE}
\end{eqnarray}
At high energies, we see that \emph{ the helicity $\pm 1/2$ gravitino tends
to the unphysical $\partial^\mu\psi_\mu$ field} and therefore we can
use the gauge-fixing condition in (\ref{gg}) to obtain a relation 
between each Fourier mode of the goldstino and the 
helicity $\pm 1/2$ gravitino. (As pointed out in \cite{Casalbuoni}, 
it is essential that both $\partial^\mu\psi_\mu$
and $\eta$  have the same poles, in order to rewrite
the equality in eq.(\ref{gg}) in terms of Fourier modes).

For arbitrary values of the $\xi$ parameter it would be necessary to 
take into account 
both solutions, with $m_+$ and with $m_-$. In this case, the solution
of the goldstino equation is
\begin{eqnarray}
\eta_p(x)=\eta_p^+(x)+\eta_p^-(x).
\end{eqnarray}
From (\ref{gg}), we see that each solution is related to the
$1/2$ helicity gravitino with different proportionality constants, 
i.e. for negative frequency solution we will have in the initial region
\begin{eqnarray}
\tilde\psi_{\pm 1/2}(\vec p)=\sum_{+,-} 
\left[-i\frac{1}{\xi}\left(1-\xi\frac{m_{+,-}^{in}}
{m_{in}}\right)P_{\pm 1/2}
+\Od\left(\frac{m_{in}}{p}\right)\right]\tilde\eta^{+,-}(\vec p).
\end{eqnarray}
In the S-matrix derivations of the ET,
the proportionality constant between
the helicity $\pm1/2$ gravitinos and the goldstinos disappears
once the external lines
of the Green functions have been removed, the momenta are on-shell, and
the tensor indices are contracted with the corresponding polarization 
vectors \cite{Casalbuoni}.
However, this is not so straightforward in the ``semiclassical''
proofs based on the generating functional formalism, either within
supergravity \cite{Casalbuoni} or even in the non-abelian context 
\cite{Cornwall}.
These ``semiclassical'' proofs are given for the clever choice $\xi=3/2$,
where $m_-=m_+$ and the proportionality constant is unique (in the
non-abelian case the choice is $\xi=1$ and the proportionality
constant is unity).

However, for our purposes, it is much more appropriate
to choose the Landau gauge. Indeed, in an arbitrary 
generalized $R_\xi$ gauge, eq. (\ref{gold0}) will be
written 
\begin{eqnarray}
i\Dbar \eta -e^{G/2}\left(G_{,\phi\phi}+G_{,\phi}^2\right)\eta
-e^{G/2}G_{,\phi}\frac{i}{2\xi\Dbar}e^{G/2}G_{,\phi}\eta=0
\label{goldxi}
\end{eqnarray}
The presence of the last term makes it very difficult
to obtain solutions even numerically. 
However, we 
get a dramatic simplification
by using the Landau gauge, 
$\xi\rightarrow \infty$, in which the last term, which is 
the most complicated, vanishes. Thus we have:
\begin{eqnarray}
i\Dbar \eta - e^{G/2}\left(G_{,\phi\phi}+G_{,\phi}^2\right)\eta=0
\label{goldstin}
\end{eqnarray}
Note that this last expression corresponds to the $m_+$ case in
(\ref{gold2}). The $m_-=0$ solution is just an artifact due to the
multiplication by $i\pabar$ and
 we do not have to consider it in the previous formulae. 
Therefore, eq.(\ref{preTE}) now reads
\begin{equation}
\tilde\psi_{\pm 1/2}(\vec p)=\left[2\,i \,P_{\pm}
+\Od\left(\frac{m_{in}}{p}\right)\right]\eta(\vec p).
\label{TE}
\end{equation}
This is the relation we were looking for.
 Note that this is an equality at
the level of fields and not for S-matrix elements. 
This result shows that 
\emph{although the helicity $\pm 1/2$ gravitinos and the goldstinos
can evolve differently} during the oscillations of the
scalar field, \emph{they approach each other 
in the asymptotic regions} (up to a constant). 
The fact that this result is
valid  only in the asymptotic regions, is sufficient  
for our purposes. Since (\ref{goldstin})
is just the standard equation of motion for a fermion field in 
curved space-time with a time-dependent mass, it is straightforward to
apply the standard techniques of particle production.

Once more we stress that 
such a result is only useful when the energy of the particles we are
producing is much larger than their masses.  In our case, 
the inflaton is in a frame where it is homogeneous, only 
depends on $t$, and oscillates with a typical frequency $m_{\phi}$.
Thus, we expect the physical momentum of the gravitinos to 
be ${\cal O}(m_{\phi})$,
so that we can use eq. (\ref{TE}) if $m_{3/2}\ll m_{\phi}$.
Usually we will evaluate $a_{out}$ right after the preheating ends  
and then $m_{in,out}\ll m_{3/2}$. 
The fact that we are in an homogeneous background is 
technically relevant 
due to the remarks about the ET and Lorentz 
invariance done in \cite{ETLorentz} in the context of
non-abelian gauge theories. Indeed the ET is not 
Lorentz invariant, since not only is
the helicity decomposition  frame-dependent,
but also terms like  $\Od (m/E)$, $\Od (m/p)$... do have very different
values depending on the reference frame. 
The fact that the inflaton field is homogeneous 
ensures that all our gravitinos are produced from 
rather similar conditions and we can apply the ET to the vast majority of them.

\section{Particle production}
Up to now our discussion has been purely classical.
 In order to interpret these solutions
in terms of particle number, we have to quantize them
\cite{Birrell,Mostepanenko}, which has already been done
in the unitary gauge \cite{Linde}. 
We are interested in the Landau gauge, $\xi\rightarrow \infty$, where
the goldstino equation of motion (\ref{goldstin}) reduces to 
that of a Majorana fermion coupled to a scalar field 
in a curved space-time,
whose quantization is also a well known problem of
 particle production  \cite{Baacke,Green}.

Let us then consider a classical solution to eq.(\ref{goldstin}) 
with helicity $l$, such
that in the past $(t\rightarrow -\infty)$, 
it behaves as a negative-energy plane-wave, i.e.:
\begin{eqnarray}
\eta^{p}_l(x)\rightarrow \frac{1}{a_{in}^{3/2}\sqrt{2\omega_{in}^+}}
e^{i\omega_{in}^+ t -i\vec p \vec x}u(\vec p, l),
\end{eqnarray}
where $a_{in}$ is the scale factor at the end of inflation. 
 In the asymptotic future
$(t\rightarrow \infty)$, because 
of the presence of the time-dependent background fields, this solution
will no longer behave as a negative-energy mode; rather, it will be a linear
superposition of positive and negative frequency modes
\begin{eqnarray}
\eta^{p}_l(x)\rightarrow \frac{1}{a_{out}^{3/2}\sqrt{2\omega_{out}^+}}
\left(\alpha^\eta_{p,l} 
e^{i\omega_{out}^+ t -i\vec p \vec x}
u(\vec p, l) +\beta^\eta_{-p,l} e^{-i\omega_{out}^+ t -i\vec p \vec x}
u^C(-\vec p, l)\right),
\end{eqnarray} 
where, 
for a given $\vec p$, we have $(\omega_{in,out}^+)^2=(m_{in,out}^+)^2+p^2$.  
The Bogolyubov coefficients satisfy
\begin{eqnarray}
\vert \alpha^\eta_{p,l} \vert^2+\vert \beta^\eta_{p,l}\vert^2=1.
\label{norm}
\end{eqnarray}
Using again our previous result in (\ref{TE}) we can identify each Fourier
mode above with the corresponding Fourier mode for a helicity $\pm 1/2$ 
gravitino (up to a constant). In particular we will find that the
Bogolyubov coefficient for the helicity $\pm 1/2$ gravitino
will be the same as that of the goldstino,  up to  $\Od(m/p)$, i.e 
$\beta^\psi_{p,l}=(1+\Od(m/p))\beta^\eta_{p,l}$.
Notice that because of the different masses of goldstinos and
physical gravitinos, the  
correction to the Bogolyubov coefficients for gravitinos can  
depend  on time as $\exp(i\Delta_{\omega} t)$, where 
$\omega_{out}^+=\omega_{out}+\Delta_{\omega}$. However, such a term
will be relevant only when $t\simeq 1/\Delta_{\omega}$, that is, much
later than the end of preheating.  
Moreover, remember that the
Bogolyubov coefficients are normalized according to (\ref{norm}) and
therefore {\em the proportionality constant is irrelevant}.

As a remark, let us note that if we had a renormalizable theory whose 
low-energy limit is supergravity, we could still 
use our estimates with the ET, irrespective of
the renormalization corrections \cite{ETrenormalizado} 
needed in the complete proof of the theorem. The reason is that
we do not need the proportionality constant to obtain the 
Bogolyubov coefficients.

Therefore the number of  gravitinos created  with helicity
$l=\pm 1/2$ and momentum $p$, $N^\psi_{p,l}$
will be  given by
\begin{eqnarray}
N^\psi_{p,l}=\left[1+\Od\left(\frac{m}{p}\right)\right]\
 \vert \beta^\eta_{p,l} \vert ^2
\label{number}
\end{eqnarray}
In conclusion, solving the equation of motion for the goldstinos
in the presence of the external backgrounds,
and using (\ref{number}), we will obtain the number of helicity $\pm
1/2$ 
gravitinos created during preheating.
In the above expression it is explicit that only the knowledge of the
solutions in the asymptotic regions is relevant to the particle
number calculation.

As an analytic check of the ET, we can compare with the unitary
gauge results of \cite{Linde}, obtained  in the global supersymmetric
limit, $\vert \phi \vert\ll  M_P$. We only have to note that,
when $\vert \phi \vert\ll  M_P$, our  goldstino
equation in the Landau gauge (\ref{goldstin}) is reduced to
\begin{eqnarray}
i\Dbar \eta - \left(\partial_\phi \partial_\phi W\right)\eta=0,
\label{susy}
\end{eqnarray}
which is the very same equation obtained in \cite{Linde} for the helicity
$\pm 1/2$ gravitinos. 
Note however, that the condition $\vert \phi \vert \ll M_P$, is not necessary
to prove the ET. Our high-energy result is valid independently of the
size of $\vert \phi \vert$.


\section{Numerical example}

As a possible application of the previous results and 
for further comparison with other works \cite{Linde}, we will study a
simple supergravity model based on the superpotential
\begin{eqnarray}
W=\sqrt{\lambda}\frac{\Phi^3}{3}.
\label{super}
\end{eqnarray}
Unfortunately, at the minimum of the 
corresponding inflaton potential, supersymmetry
is restored. We will then assume that (\ref{super}) is
only valid far from the minimum and that close to $\phi=0$ 
it is modified to satisfy the
assumptions needed to apply the ET (see paragraph before eq.(\ref{eomgravitino})). 
For this superpotential, the 
effective gravitino mass oscillations are not damped in time. Still, we can 
apply the ET if we take
$\phi(t)=0$ for $t\leq 0$ and $t\geq nT$, where $n$ is an integer
(in our example $n=6$) 
and $T$ is the inflaton oscillation period.
Despite these problems, we consider this model
useful as a numerical illustration of the ET.
Hence, we have taken the initial amplitude of
the inflaton oscillation to be $\tilde\phi_0\simeq 0.2 M_P$
(where $\tilde \phi=\sqrt{2}\re \phi$ is the canonically normalized
inflaton), which
implies that the effective goldstino mass  is
oscillating with amplitude $m_G\simeq \sqrt{2\lambda} M$. 
With these initial and final conditions we have calculated numerically
the number of goldstinos produced from eq. (\ref{goldstin}),
using the standard
results for the production of fermions obtained in \cite{Baacke}.
We thus look for solutions of the goldstino equation with momentum $p$
and helicity $l$ of the form
\begin{eqnarray}
\eta^{pl}(x)=a^{-3/2}(t)e^{i\vec p \cdot \vec x} U^{\vec p l}(t),
\end{eqnarray}
with 
\begin{eqnarray}
U^{\vec p l}(t)=\frac{1}{\sqrt{\omega_{in}+m_{in}}}\left[
i\gamma^0\partial_0
-\vec p\cdot \vec\gamma+ a(t)\left(
e^{G/2}\left(G_{,\phi\phi}+G_{,\phi}^2\right)
\right)\right]f_{pl}(t)u(\vec p, s).
\end{eqnarray}
Using the above ansatz, we can write the equation as follows:
\begin{eqnarray}
\left[\frac{d^2}{d\tilde t^2}+\kappa^2+\frac{i}{\sqrt{\lambda}}
\frac{d}{d \tilde t}\left(
be^{G/2}\left(G_{,\phi\phi}+G_{,\phi}^2\right)\right)+
\frac{b^2}{\lambda} e^{G}
\left(G_{,\phi\phi}+G_{,\phi}^2\right)^2\right]f_{\kappa l}(\tilde t)=0
\label{master2}
\end{eqnarray}
with $\kappa=p/(a_{in}\sqrt{\lambda})$ and $\tilde t=a_{in}\sqrt{\lambda}t$ 
and the new scale factor is defined as 
$b(\tilde t)=a(\tilde t)/a_{in}$. The initial conditions are 
$f_{\kappa l}(0)=1$ and $\dot f_{\kappa l}(0)=-i\kappa$. In
particular, for the goldstino occupation number we find
\begin{eqnarray}
N^\eta_{\kappa l}(nT)=\frac{1}{4\kappa}\left(2 \kappa
+i[\dot f^*_{\kappa l}(nT) 
f_{\kappa l}(nT)- f^*_{\kappa l}(nT)
\dot f_{\kappa l}(nT)]\right)
\label{occupation}
\end{eqnarray}
Using eq.(\ref{number}), we get the occupation number of 
helicity $\pm 1/2$ gravitinos directly from (\ref{occupation}),
whereas for those with helicity $\pm 3/2$
we use eq. (\ref{g32}) following the
same steps as before \cite{Maroto}. In Fig. 1, we have plotted both 
spectra and
we see that for this particular model, the production of helicity
$\pm 3/2$ gravitinos is suppressed by two to three orders of magnitude 
with respect to the $\pm 1/2$ gravitinos. Note that 
$N_{kl}$ depends on the number of oscillations 
 and on the inflaton initial conditions. Thus,  we cannot
 make a comparison for the whole spectrum with the unitary gauge
estimations obtained in the first references in \cite{Linde} and the 
complete numerical analysis of the last reference
in \cite{Linde} (made without the ET, although they have checked that the 
formulas hold in one particular case). Nevertheless, 
the ET tells us that, since the mass of the goldstino will
be less than $a(t)\sqrt{\lambda}$, both results should agree for 
$p\gg a(nT)\sqrt{\lambda}$, i.e.  with
the above definitions $\kappa\gg b(nT)$ (in our example $b(nT)\simeq
12$), irrespectively of the initial conditions. Indeed, the
orders of magnitude are in good agreement with the previous works. 


\begin{figure}[htbp]
\begin{center}
\hspace*{-1.4cm}
\hbox{\psfig{file=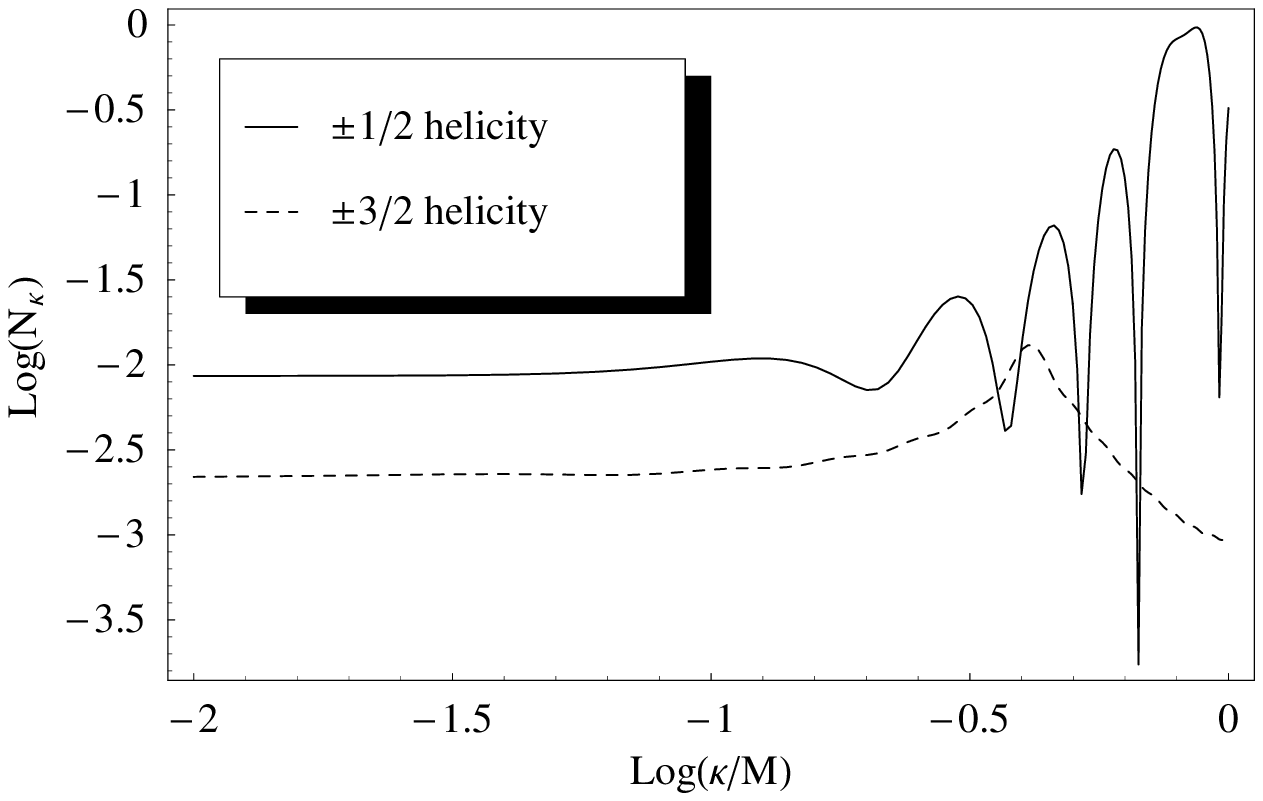,width=9cm}}
\hbox{\psfig{file=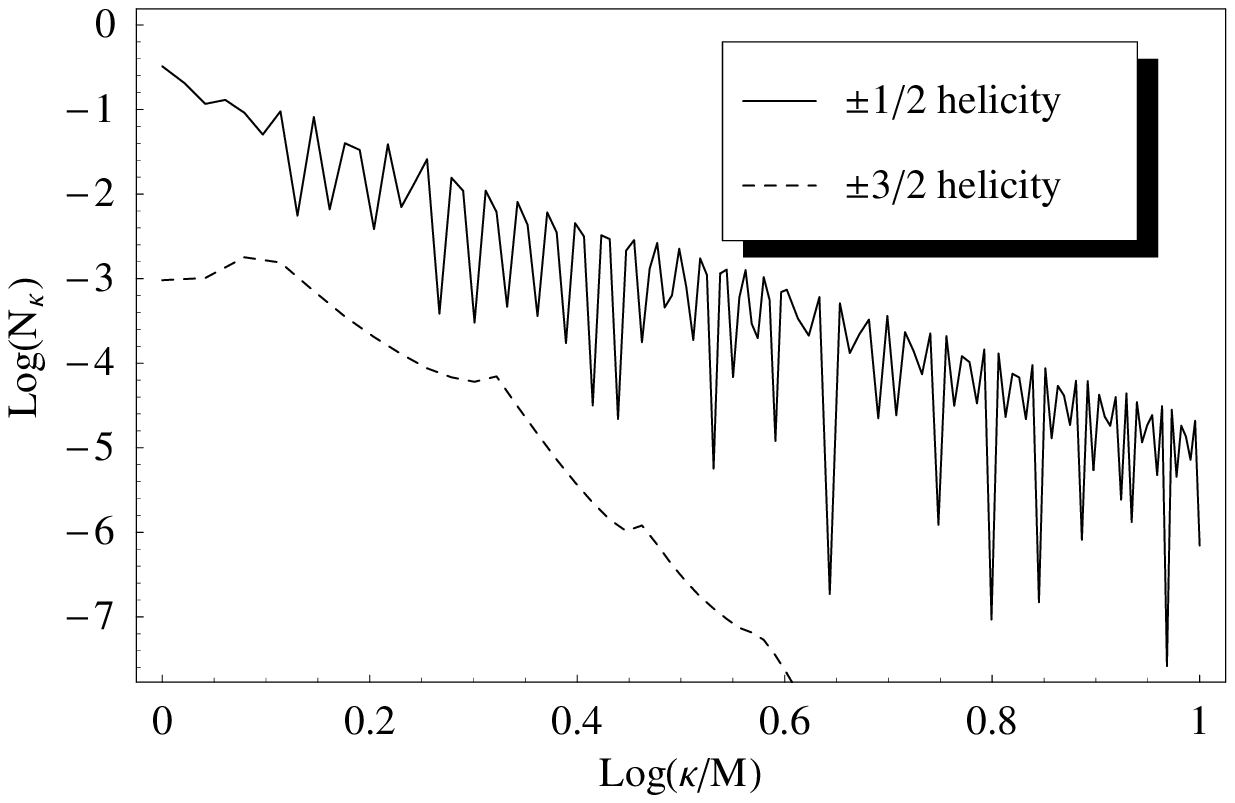,width=9cm}}
\caption{Spectrum of  helicity $\pm 3/2$ and helicity $\pm 1/2$
gravitinos for different ranges of momentum. The helicity $\pm 1/2$ 
production has been obtained using the
ET.}
\label{fig:result}
\end{center}
\end{figure}
       

However, there 
are some other models in which the application of the 
ET is somewhat restricted, such as a 
pure quadratic superpotential $W=m_{\phi}\Phi^2$. In this case, at the minimum
of the potential, supersymmetry is restored and therefore the
definition of helicity $\pm 1/2$ gravitinos is meaningless.
Ignoring this problem, although the 
amplitude of the inflaton oscillations is damped, 
the goldstino mass term, which contains the second derivative of the 
superpotential, 
tends to a constant of ${\cal O}(m_{\phi})$, typically much larger
than $m_{3/2}$. Then, the ET will be useful to calculate
only that portion of the spectrum with energy much higher than the inflaton 
mass.

Finally, there is an additional condition to 
obtain  reliable  predictions. Since
gravitinos are produced in large numbers, with vastly 
different energies, we have to estimate 
what fraction of them does not satisfy
the ET conditions. For that purpose we define the number density of
gravitinos produced with both helicities as
\begin{eqnarray}
n(t)=\frac{1}{\pi^2 a^3(t)}\int_0^{\infty} N_{p,l}p^2 dp.
\end{eqnarray}
Then, for those gravitinos with  momenta lower than 
their mass, i.e. $p^2<a_{out}^2
m_{3/2}^2$, the ET does not apply. Thus 
the number density of ``excluded'' gravitinos is
\begin{eqnarray}
n(t)_{exc}\leq\frac{1}{3\pi^2 a^3(t)}a_{out}^3 m_{3/2}^3
\end{eqnarray}
In addition, we can estimate the total number density of
produced gravitinos as:
\begin{eqnarray}
n(t)_{tot}\simeq\frac{1}{3\pi^2 a^3(t)}a_{in}^3 m_{\phi}^3
\end{eqnarray}
Hence we obtain that the fraction of gravitinos 
that do not satisfy the ET applicability conditions is
\begin{eqnarray}
\frac{n_{exc}}{n_{tot}}\leq \left( \frac{a_{out} m_{3/2}}{a_{in}
  m_{\phi}}\right)^3
\end{eqnarray}
Accordingly, the additional ET applicability condition  is:
$a_{out}m_{3/2}<<a_{in}m_{\phi}$.
We see that the result depends on the duration of the preheating era
and the ratio of gravitino and inflaton masses. 
Typically, the production takes place in a few inflaton oscillations,
which implies that the scale factor only grows by a few orders of
magnitude, not enough to overcome the mass difference. 
Therefore, in these models, the ET safely describes the production
of the vast majority of helicity $\pm 1/2$ gravitinos.

\section{Conclusions}
We have studied the production of helicity $\pm 1/2$ gravitinos using the
equivalence of goldstinos and gravitinos at high energies. 
We have shown that in the $R_\xi$ gauges, 
where the goldstino and gravitino equations of motion decouple,
the classical solutions of the goldstino equation are
proportional to those of the helicity $\pm 1/2$ gravitinos
in the asymptotic static regions.
This result is 
sufficient to relate the production of goldstinos to the production of
helicity $\pm 1/2$ gravitinos. Furthermore, we have shown how in the
Landau gauge, the equation
of motion for the goldstino is considerably simpler.
As a check, we have compared our results  
with previous ones obtained in the unitary gauge
and we have found  good agreement in the equivalence theorem
applicability regions. 

We have clearly identified the ET applicability conditions in the
context of gravitino production:
i) The frequency of the inflaton field  oscillations should be 
larger than the gravitino mass, $m_{\phi}\gg m_{3/2}$. 
If we are interested in the pure gravitational production,
then one should also require $H\gg m_{3/2}$. These
conditions ensure that the typical energy of the 
produced particles will be larger than their masses 
ii) The sources should
vanish asymptotically, which implies that the space-time curvature
should decrease with time and also that the amplitude of the inflaton
oscillations should be damped.  
iii) The ET calculations are only useful if most
of the gravitinos have a large enough energy, which
requires $a_{out}m_{3/2}\ll a_{in}m_\phi$.

Concerning the potential applications of our results, we have shown
that cubic superpotentials, with a slight modification 
in their form, together with appropriate initial conditions, 
could satisfy the above applicability requirements. 
More elaborated models would require to extend the present proof to
more than one chiral multiplet and non-minimal supergravity.
Further work along this line is in progress.

\section*{Acknowledgements}
A.L.M.  wishes to acknowledge very useful
discussions with A. Linde, R. Kallosh and A. Mazumdar.
J.R.P. thanks the CERN Theory Division, 
where part of this work was carried out, for their hospitality.
This work has been partially supported
by the Spanish Ministerio de Educaci\'on y Ciencia
under project CICYT AEN97-1693. A.L.M. also
acknowledges the Spanish Ministerio de Educaci\'on y Ciencia 
for a Post-Doctoral Fellowship.

\newpage

\thebibliography{references} 

\bibitem{Nilles} H.P. Nilles, {\it Phys. Rep.} {\bf 110}, 1 (1984) 
\bibitem{Bailin} D. Bailin and A. Love, {\it Supersymmetric Gauge Field
Theory and String Theory}, IOP, Bristol, (1994)
\bibitem{Kolb} E.W. Kolb and M.S Turner, {\it The Early Universe} 
(Addison-Wesley, New York, 1990)
\bibitem{LiEll} J. Ellis, A. Linde and D. Nanopoulos, {\it Phys. Lett.}
{\bf 118B}, 59 (1982)
\bibitem{Sarkarrep} S. Sarkar, {\it Rep. Prog. Phys.} {\bf 59}, 1493 (1996)
\bibitem{Ellis} J. Ellis, J.E. Kim and D.V. Nanopoulos, {\it Phys. Lett.}
{\bf 145B}, 181 (1984)
\bibitem{LindeB} L. Kofman, A. D. Linde and A.A. Starobinsky,
{\it Phys. Rev. Lett.} {\bf 73}, 3195 (1994);
L.Kofman, A.D. Linde and A. A. Starobinsky, {\it Phys. Rev.} {\bf D
  56}, 3258 (1997); J.H. Traschen, R.H. Brandenberger, {\it
  Phys. Rev.} {\bf D42}, 2491 (1990); Y. Shtanov, J. Traschen and 
R. Brandenberger, {\it Phys. Rev.} {\bf D51}, 5438 (1995)
\bibitem{Baacke} J. Baacke, K. Heitmann and C. Patzold, {\it Phys. Rev.}
 {\bf D58}, 125013 (1998)
\bibitem{Green} P.B. Greene and L. Kofman, {\it Phys. Lett.} {\bf B448}, 6
(1999)
\bibitem{Maroto} A.L. Maroto and A. Mazumdar, {\it Phys. Rev. Lett.} 
{\bf 84} (2000) 1655.
\bibitem{Valencia}  A.L. Maroto,{\it   Nucl. Phys. B (Proc. Suppl.)}
  {\bf 81}, 351 (2000) hep-ph/9906388
\bibitem{Lemoine} M. Lemoine, {\it Phys. Rev.} {\bf D60}, 103522 (1999)
\bibitem{Linde} R. Kallosh, L. Kofman, A. Linde and A. Van Proeyen,
hep-th/9907124, G.F. Giudice, A. Riotto and I. Tkachev, 
{\it JHEP} (1999) 9908:009, G.F. Giudice, A. Riotto and I. Tkachev, hep-ph/9911302  
\bibitem{varios} D.H. Lyth, D. Roberts and M. Smith, {\it Phys. Rev.} {\bf D57}
7120 (1998); D.H. Lyth, hep-ph/9909387 and hep-ph/9911257
\bibitem{Fayet} P. Fayet, \textit{Phys. Lett} {\bf B70}, (1977) 461
\bibitem{Casalbuoni} R. Casalbuoni, S. De Curtis, D. Dominici,
  F. Feruglio and R. Gatto, {\it Phys. Lett.} {\bf B215}, 313 (1988)
and  {\it Phys. Rev.} {\bf D39}, 2281 (1989).
\bibitem{Cornwall} J.M. Cornwall, D.N. Levin and G. Tiktopoulos, {\it 
Phys. Rev.}
{\bf D10}, 1145 (1974)
\bibitem{ETgauge}
C.E. Vayonakis, {\it Lett. Nuovo. Cim.} {\bf 19}, 383 (1976)\\
B.W. Lee, C. Quigg and H. Thacker, {\it Phys. Rev.} {\bf D16}, 1519 (1977)
\bibitem{Chanowitz} M.S. Chanowitz and M.K. Gaillard, 
{\it Nucl. Phys.} {\bf B261}, 379 (1985)    
\bibitem{ETrenormalizado} Y.P.Yao and C.P.Yuan, {\it Phys. Rev.} {\bf
    D38}, 2237 (1988)\\
J.Bagger and C.Schmidt,  {\it Phys. Rev.} {\bf D41}, 2237 (1990)\\
H.J.He, Y.P. Kuang and X. Li, {\it Phys. Rev. Lett.} {\bf 69}, 2619 (1992).
\bibitem{ETLorentz} H.J.He, Y.P. Kuang and X. Li, 
{\it Phys. Rev.} {\bf D49}, 4842 (1994).
\bibitem{Julia} E. Cremmer, B. Julia, J. Scherk, S. Ferrara,
  L. Girardello and P. van Nieuwenhuizen, {\it Nucl. Phys.} {\bf 147},
  105 (1979)
\bibitem{Holman} P. Holman, P. Ramond and G.G. Ross, {\it Phys. Lett.}
{\bf 137B}, 343 (1984)
\bibitem{Ross} G.G. Ross and S. Sarkar, {\it Nucl. Phys.} {\bf B461},
  597 (1996)
\bibitem{Baulieu} L. Baulieu, A. Georges and S. Ouvry, {\it
    Nucl. Phys.} {\bf B273}, 366 (1986)
\bibitem{Birrell} N.D. Birrell and P.C.W.
Davies {\it Quantum Fields in Curved Space}, Cambridge University Press 
(1982)

\bibitem{Moroi} T. Moroi, PhD Thesis (Tohoku University) (1995), 
hep-ph/9503210 
\bibitem{Mostepanenko} V.M. Mostepanenko and V.M Frolov, 
{\it Sov. J. Nucl. Phys.}
{\bf 19} 451 (1974); A.A. Grib, S.G. Mamayev and V.M. Mostepanenko {
\it Vacuum Quantum Effects in Strong Fields}, Friedmann Laboratory 
Publishing, St. Petersburg (1994)
\end{document}